\documentclass{sig-alternate}

\usepackage{amsthm,amssymb,amsmath}
\usepackage{graphics}
\usepackage{tikz}
\usepackage[plainpages=false,pdfpagelabels,colorlinks=true,citecolor=blue,hypertexnames=false]{hyperref}
\usepackage{color}

\def\eatspace#1{#1}
\def\step#1#2{\par\kern1pt\dimen44=#2em\advance\dimen44 1.67em\hangindent\dimen44\hangafter=1\noindent\rlap{\small#1}\kern\dimen44\relax\eatspace}

\newtheorem{theorem}{Theorem}

\newtheorem{defi}[theorem]{Definition}
\newtheorem{example}[theorem]{Example}

\def\qed{\quad\rule{1ex}{1ex}}

\let\set\mathbb

\def\lcm{\operatorname{lcm}}

\boilerplate={}
\copyrightetc={}

\begin{document}

\title{Factorization of C-finite Sequences}

\numberofauthors{2}

\author{%
 \alignauthor
 \leavevmode\mathstrut Manuel Kauers\thanks{partially supported by FWF grants F50-04 and Y464-N18.}\\[\medskipamount]
  \affaddr{\leavevmode\mathstrut Institute for Algebra}\\
  \affaddr{\leavevmode\mathstrut Johannes Kepler University}\\
  \affaddr{\leavevmode\mathstrut 4040 Linz, Austria}\\
  \affaddr{\leavevmode\mathstrut manuel.kauers@jku.at}
 \and
 \mathstrut Doron Zeilberger\\[\medskipamount]
  \affaddr{\mathstrut Department of Mathematics}\\
  \affaddr{\mathstrut Rutgers University}\\
  \affaddr{\mathstrut New Brunswick, NJ, USA}\\
  \affaddr{\mathstrut zeilberg@math.rutgers.edu}
}

\maketitle
\begin{abstract}
  We discuss how to decide whether a given C-finite sequence
  can be written nontrivially as a product of two other
  C-finite sequences. 
\end{abstract}








\section{Introduction}

It is well known that when $(a_n)_{n=0}^\infty$ and $(b_n)_{n=0}^\infty$ are two
sequences that satisfy some linear recurrences with constant coefficients, then
the product sequence $(a_nb_n)_{n=0}^\infty$ also satisfies such a recurrence.
Sequences satisfying linear recurrences with constant coefficients are called
C-finite~\cite{zeilberger90,kauers10j,zeilberger13}, and the fact just refered to is one of several closure
properties that this class of sequences enjoys.  In this paper, we will consider
the inverse problem: given a C-finite sequence $(c_n)_{n=0}^\infty$, can we
write it in a nontrivial way as the product of two other C-finite sequences?
This question is of interest in its own right, but it is also useful in some
applications in combinatorics.  For example, the celebrated solution by
Kasteleyn, and Temperley-Fisher, of the dimer problem~\cite{fisher61,kasteleyn61} as well as the even
more celebrated Onsager solution of the two-dimensional Ising model~\cite{onsager44}
can be (re)discovered using an algorithm for factorization of C-finite
sequences.

A C-finite sequence is uniquely determined by a recurrence and a choice
of sufficiently many initial values. The prototypical example of a C-finite
sequence is the Fibonacci sequence $(F_n)_{n=0}^\infty$ defined by
\[
 F_{n+2}-F_{n+1}-F_n=0,\qquad F_0=0,F_1=1.
\]
Whether a C-finite sequence $(c_n)_{n=0}^\infty$ admits a factorization depends
in general on both the recurrence as well as the initial values. For example,
the sequence $(3^n+4^n+6^n+8^n)_{n=0}^\infty$, which satisfies the recurrence
\[
 c_{n+4}-21c_{n+3}+158c_{n+2}-504c_{n+1}+576c_n=0,
\]
can be factored as $3^n+4^n+6^n+8^n=(1+2^n)(3^n+4^n)$, while the sequence
$3^n+4^n+6^n-8^n$, which satisfies the same recurrence, cannot be factored.

We shall consider a variant of the factorization problem that does not depend
on initial values but only on the recurrence equations. Linear recurrences
may be viewed as polynomials $p=p_0+p_1x+\cdots+p_dx^d\in k[x]$ acting on
sequences $(a_n)_{n=0}^\infty$ via
\[
  p\cdot(a_n)_{n=0}^\infty:=(p_0a_n+p_1a_{n+1}+\cdots+p_da_{n+d})_{n=0}^\infty.
\]
For every fixed~$p\in k[x]$, denote by $V(p)$ the set of all sequences
$(a_n)_{n=0}^\infty$ with $p\cdot(a_n)_{n=0}^\infty=(0)_{n=0}^\infty$, i.e., the solution
space of the recurrence equation encoded by~$p$.
This is a
vector space of dimension~$\deg(p)$. For any two operators $p,q\in k[x]\setminus\{0\}$ there
exists a unique monic polynomial $r\in k[x]$ such that $V(r)$ is
vector space generated by all sequences $(a_nb_n)_{n=0}^\infty$
with $(a_n)_{n=0}^\infty\in V(p)$ and $(b_n)_{n=0}^\infty\in V(q)$,
i.e., $V(r)=V(p)\otimes V(q)$. We write $r=p\otimes q$. 

Our problem shall be to decide, for a given monic polynomial $r\in k[x]$,
whether there exist $p,q\in k[x]$ such that $r=p\otimes q$. In principle, it is
known how to do this. Singer~\cite{singer85a} gives a general algorithm for the
analogous problem for linear differential operators with rational function
coefficients, the problem is further discussed in~\cite{hessinger97}. Because of their
high cost, these algorithms are mainly of theoretical interest. For the special
case of differential operators of order 3 or~4 (still with rational function
coefficients), van Hoeij~\cite{hoeij07,hoeij02a} combines several observations to 
algorithms which handle these cases efficiently. For the recurrence case,
Cha~\cite{cha14} gives an algorithm for operators of order~3 with rational function
coefficients. An algorithm for the case of constant coefficients and arbitrary
order was recently sketched by the second author~\cite{zeilberger13}. This description
however only considers the ``generic case''. The present paper is a
continuation of this work in which we give a complete algorithm which also
handles ``degenerate'' cases. Our algorithm is efficient in the sense that it
does not require any Gr\"obner basis computation, but inefficient in the sense
that it requires a search that may take exponential time in the worst
case. 

\section{Preliminaries}

To fix notation, let us recall the basic facts about C-finite sequences.
Let $k$ be an algebraically closed field. 

\begin{defi}
  \begin{enumerate}
  \item A sequence $(a_n)_{n=0}^\infty$ is called \emph{C-finite,} if there exist
    $p_0,\dots,p_d\in k$ with $p_0\neq0\neq p_d$ such that for all $n\in\set N$ we have
    $p_0a_n + \cdots p_da_{n+d}=0$.
  \item In this case, the polynomial $p=p_0+p_1x+\cdots+p_dx^d$ is called a
    \emph{characteristic polynomial} for $(a_n)_{n=0}^\infty$.
  \item For $p\in k[x]$, the set $V(p)$ denotes the set of all C-finite sequences
    whose characteristic polynomial is~$p$. It is called the \emph{solution space}
    of~$p$.
  \end{enumerate}  
\end{defi}

\begin{theorem}\cite{stanley99,kauers10j}\label{thm:main}
  Let
  $p=(x-\phi_1)^{e_1}\cdots(x-\phi_m)^{e_m}\in k[x]$
  for pairwise distinct $\phi_1,\dots,\phi_m\in k\setminus\{0\}$. Then $V(p)$
  is the $k$-vector space generated by the sequences
  \begin{alignat*}1
    &\phi_1^n,\ \dots,\ n^{e_1-1}\phi_1^n,\\
    &\phi_2^n,\ \dots,\ n^{e_2-1}\phi_2^n,\\
    &\dots\dots\dots,\\
    &\phi_m^n,\ \dots,\ n^{e_m-1}\phi_m^n.
  \end{alignat*}
\end{theorem}

It is an immediate consequence of this theorem that for any two polynomials $p,q\in k[x]$
we have $V(\gcd(p,q))=V(p)\cap V(q)$ and $V(\lcm(p,q))=V(p)+V(q)$. The latter says in
particular that when $(a_n)_{n=0}^\infty$ and $(b_n)_{n=0}^\infty$ are C-finite, then so is
their sum $(a_n+b_n)_{n=0}^\infty$. A similar result holds for the product: write
$p=\prod_{i=1}^m (x-\phi_i)^{e_i}$ and $q=\prod_{j=1}^\ell (x-\psi_j)^{\epsilon_j}$ and define
\begin{equation}\label{eq:1}
  r:=p\otimes q:=\lcm_{i=1}^m\lcm_{j=1}^\ell (x-\phi_i\psi_j)^{e_i+\epsilon_j-1}.
\end{equation}
Then $r$ is a characteristic polynomial for the product sequence $(a_nb_n)_{n=0}^\infty$.
Note that $\deg(p)+\deg(q)\leq\deg(r)\leq\deg(p)\deg(q)$ for every $p,q\in k[x]$.
Note also that $p\otimes q=q\otimes p$ for every $p,q\in k[x]$.

Our goal is to recover $p$ and $q$ from a given~$r$.
The problem is thus to decide whether the roots of a given polynomial $r$ are precisely
the pairwise products of the roots of two other polynomials $p$ and~$q$.
Besides the interpretation as a factorization of C-finite sequences, this problem can
also be viewed as factorization of algebraic numbers: given some algebraic number~$\alpha$,
specified by its minimal polynomial~$r$, can we write $\alpha=\beta\gamma$ where $\beta,\gamma$
are some other algebraic numbers with respective minimal polynomials $p$ and~$q$. 

Trivial decompositions are easy to find: For each $r$ we obviously have $r=r\otimes(x-1)$.
Moreover, for every nonzero $\phi$ we have $(x-\phi)\otimes(x-\phi^{-1})=(x-1)$, so we
can ``decompose'' $r$ into $r\otimes(x-\phi)$ and $x-\phi^{-1}$.
In order for a decomposition $r=p\otimes q$ to be interesting, we have to require that
both $p$ and $q$ have at least degree~2.

Even so, a factorization is in general not unique. Obviously, if $r=p\otimes q$ is a factorization,
then for any nonzero~$\phi$ also $r=\bigl(p\otimes(x-\phi)\bigr)\otimes\bigl((x-\phi^{-1})\otimes q\bigr)$.
Translated to sequences, this ambiguity corresponds to the facts that for every $\phi\neq0$, both
$(\phi^{n})_{n=0}^\infty$ and $(\phi^{-n})_{n=0}^\infty$ are C-finite, and that a sequence $(a_n)_{n=0}^\infty$ is C-finite
iff for all $\phi\neq0$ the sequence $(a_n\phi^n)_{n=0}^\infty$ is C-finite. 
But there is even more non-uniqueness: the polynomial
\[
  r=(x-2)(x+2)(x-3)(x+3)
\]
admits the two distinct factorizations
\begin{alignat*}1
  r&=(x-1)(x+1)\otimes(x-2)(x+3)\\
   &=(x-1)(x+1)\otimes(x-2)(x-3)
\end{alignat*}
which cannot be obtained from one another by introducing factors $(x-\phi)$ and
$(x-\phi^{-1})$. Our goal will be to compute a finite list of factorizations
from which all others can be obtained by introducing factors
$(x-\phi)\otimes(x-\phi^{-1})$.

There is a naive but very expensive algorithm which does this job when $r$ is
squarefree: For some choice $n,m$ of degrees, make an ansatz
$p=(x-\phi_1)\cdots(x-\phi_n)$ and $q=(x-\psi_1)\cdots(x-\psi_m)$ with variables
$\phi_1,\dots,\phi_n,\psi_1,\dots,\psi_m$.  Equate the coefficients of $r -
\prod_{i=1}^n\prod_{j=1}^m (x-\phi_i\psi_j)$ with respect to $x$ to zero and
solve the resulting system of algebraic equations for
$\phi_1,\dots,\phi_n,\psi_1,\dots,\psi_m$.  After trying all possible degree
combinations $n\geq m\geq2$ with $n+m\leq\deg(r)\leq nm$, either a decomposition
has been found, or there is none.

\section{The Generic Case}

Typically, when $p$ and $q$ are square-free polynomials and $\phi_1,\dots,\phi_n\neq0$
are the roots of $p$ and $\psi_1,\dots,\psi_m\neq0$ are the roots of~$q$, then the
products $\phi_i\psi_j$ for $i=1,\dots,n$, $j=1,\dots,m$ will all be pairwise distinct.
In this case, $r=p\otimes q$ will have exactly $nm$ roots, and the factorization
problem consists in recovering $\phi_1,\dots,\phi_n$ and $\psi_1,\dots,\psi_m$
from the (known) roots $\rho_1,\dots,\rho_{nm}$ of~$r$.

As observed in~\cite{zeilberger13}, a necessary condition for $r$ to admit a factorization
into two polynomials of respective degrees $n$ and $m$ is then that there is a
bijection $\pi\colon\{1,\dots,n\}\times\{1,\dots,m\}\to\{1,\dots,nm\}$ such that for
all $j_1,j_2$ we have
\[
\frac{\rho_{\pi(1,j_1)}}{\rho_{\pi(1,j_2)}}=\frac{\rho_{\pi(2,j_1)}}{\rho_{\pi(2,j_2)}}=\cdots=\frac{\rho_{\pi(n,j_1)}}{\rho_{\pi(n,j_2)}}
\]
and for all $i_1,i_2$ we have
\[
\frac{\rho_{\pi(i_1,1)}}{\rho_{\pi(i_2,1)}}=\frac{\rho_{\pi(i_1,2)}}{\rho_{\pi(i_2,2)}}=\cdots=\frac{\rho_{\pi(i_1,m)}}{\rho_{\pi(i_2,m)}}.
\]
The explanation is simply that when a factorization exists, then the roots $\rho_\ell$ of~$r$
are precisely the products~$\phi_i\psi_j$, and if we define $\pi$ so that it
maps each pair $(i,j)$ to the corresponding root index~$\ell$,
then the quotients
\[
\frac{\rho_{\pi(i,j_1)}}{\rho_{\pi(i,j_2)}}=
\frac{\phi_i\psi_{j_1}}{\phi_i\psi_{j_2}}=
\frac{\psi_{j_1}}{\psi_{j_2}}
\]
do not depend on $i$ and the quotients
\[
\frac{\rho_{\pi(i_1,j)}}{\rho_{\pi(i_2,j)}}=
\frac{\phi_{i_1}\psi_j}{\phi_{i_2}\psi_j}=
\frac{\phi_{i_1}}{\phi_{i_2}}
\]
do not depend on~$j$.

In fact, the existence of such a bijection $\pi$ is also sufficient for the
existence of a factorization:
choose $\phi_1\neq0$ arbitrarily and set $\psi_1:=\rho_{\pi(1,1)}/\phi_1$ and 
\[\phi_i:=\phi_1\frac{\rho_{\pi(i,1)}}{\rho_{\pi(1,1)}}\qquad(i=2,\dots,n)\]
and
\[\psi_j:=\psi_1\frac{\rho_{\pi(1,j)}}{\rho_{\pi(1,1)}}\qquad(j=2,\dots,m).\]
Then we have $\rho_{\pi(i,j)}=\phi_i\psi_j$ for all $i,j$, and therefore 
for $p=(x-\phi_1)\cdots(x-\phi_n)$ and $q=(x-\psi_1)\cdots(x-\psi_m)$
we have $r=p\otimes q$.
Note that $p$ and $q$ are squarefree, because if we have, say, $\phi_{i_1}=\phi_{i_2}$
for some $i_1, i_2$, and then $\rho_{\pi(i_1,1)}=\rho_{\pi(i_2,1)}$,
and then $\pi(i_1,1)=\pi(i_2,1)$, then $i_1=i_2$.

\begin{example}
  \begin{enumerate}
  \item Consider $r=(x-4)(x-6)(x+6)(x+9)$, i.e., $\rho_1=4$, $\rho_2=6$, $\rho_3=-6$, $\rho_4=-9$.
    A possible choice for $\pi\colon\{1,2\}\times\{1,2\}\to\{1,2,3,4\}$ is given by the table
    \begin{center}
      \begin{tabular}{c|cc}
        $\pi$ & $1$ & $2$ \\\hline
        $1$ & $1$ & $2$ \\
        $2$ & $3$ & $4$
      \end{tabular}
    \end{center}
    (to be read like, e.g., $\pi(2,1)=3$), because
    \[
     \frac{\rho_{\pi(1,2)}}{\rho_{\pi(1,1)}} =
     \frac{\rho_2}{\rho_1} =
     \frac{6}{4} =
     \frac{-9}{-6} = 
     \frac{\rho_4}{\rho_3} =     
     \frac{\rho_{\pi(2,2)}}{\rho_{\pi(2,1)}}
    \]
    and
    \[
     \frac{\rho_{\pi(2,1)}}{\rho_{\pi(1,1)}} =
     \frac{\rho_3}{\rho_1} =
     \frac{-6}{4} =
     \frac{-9}{6} =
     \frac{\rho_4}{\rho_2} =
     \frac{\rho_{\pi(2,2)}}{\rho_{\pi(1,2)}}.
    \]
    Take $\phi_1=15$ (for no particular reason), $\psi_1=\frac4{15}$,
    $\phi_2=15\ \frac{6}{4}=\frac{45}2$, 
    $\psi_2=\frac4{15}\frac{(-6)}{4}=-\frac25$.
    Then
    \begin{alignat*}1
      &(x-15)(x-\tfrac{45}2)\otimes(x-\tfrac4{15})(x+\tfrac25)\\
      &=(x-15\tfrac{4}{15})(x+15\tfrac25)(x-\tfrac{45}2\tfrac4{15})(x+\tfrac{45}2\tfrac25)\\
      &=(x-4)(x+6)(x-6)(x+9),
    \end{alignat*}
    as required.

    In this example, no other factorizations exist except for those that are
    obtained by replacing $p$ and $q$ by $p\otimes(x-\xi)$ and
    $(x-\xi^{-1})\otimes q$ for some $\xi\neq0$. This degree of freedom is
    reflected by the arbitrary choice of~$\phi_1$.    
  \item The polynomial $(x-1)(x-2)(x-3)(x-4)$ cannot be written as
    $p\otimes q$ for two quadratic polynomials $p$ and~$q$, because
    $\frac12\neq\frac34$,
    $\frac12\neq\frac43$,
    $\frac13\neq\frac24$,
    $\frac13\neq\frac42$,
    $\frac14\neq\frac23$,
    $\frac14\neq\frac32$.
  \item Consider $r=(x-2)(x+2)(x-3)(x+3)$, i.e., $\rho_1=2$, $\rho_2=-2$, $\rho_3=3$, $\rho_4=-3$.
    We have seen that in this case there are two distinct factorizations.
    They correspond to the two bijections
    $\pi,\pi'\colon\{1,2\}\times\{1,2\}\to\{1,2,3,4\}$ defined via
    \begin{center}
      \begin{tabular}{c|cccc}
               & $(1,1)$ & $(1,2)$ & $(2,1)$ & $(2,2)$ \\\hline
        $\pi$  &  $1$    &  $2$     &  $3$   & $4$ \\
        $\pi'$ &  $1$    &  $2$     &  $4$   & $3$
      \end{tabular}
    \end{center}
  \end{enumerate}
\end{example}

\section{Product Clashes}

Again let $p,q\in k[x]$ be two square-free polynomials, and write
$\phi_1,\dots,\phi_n$ for the roots of $p$ and $\psi_1,\dots,\psi_m$ for the
roots of~$q$. Generically, the degree of $p\otimes q$ is equal to
$\deg(p)\deg(q)$. It cannot be larger than this, and it is smaller if and only
if there are two index pairs $(i,j)\neq(i',j')$ with
$\phi_i\psi_j=\phi_{i'}\psi_{j'}$. In this case, we say that $p$ and $q$ have a
product clash. Recall from equation~\eqref{eq:1} that $p\otimes q$ is formed as
the least common multiple of the factors $x-\phi_i\psi_j$, not as their product.

Product clashes appear naturally in the computation of $p\otimes p$.
For example, for $p=(x-\phi_1)(x-\phi_2)$ we have
\begin{alignat*}1
  p\otimes p
  &=\lcm(x-\phi_1\phi_1,x-\phi_1\phi_2,x-\phi_2\phi_1,x-\phi_2\phi_2)\\
  &=(x-\phi_1\phi_1)(x-\phi_1\phi_2)(x-\phi_2\phi_2),
\end{alignat*}
because $\phi_1\phi_2=\phi_2\phi_1$ is a clash. More generally, if $p$ is a square-free
polynomial of degree~$d\geq2$, then $\deg(p\otimes p)\leq\frac12d(d+1)<d^2$.

As an example that does not come from a product of the form $p\otimes p$,
consider $p=(x-1)(x-2)(x-4)$ and $q=(x-\tfrac12)(x-\tfrac14)$. Here we have
the clashes $1\cdot\frac12=2\cdot\frac14$ and $2\cdot\frac12=4\cdot\frac14$,
so that $p\otimes q=(x-\tfrac12)(x-\tfrac14)(x-1)(x-2)$ only has degree~4.

In order to include product clashes into the framework of the previous section,
we need to relax the requirement that $\pi$ be injective. We still want it to
be surjective, because every root of $r$ must be produced by the product $\phi\psi$
of some root $\phi$ of $p$ and some root $\psi$ of~$q$. If the $\phi_i$ and the
$\psi_j$ are defined according to the formulas above, it can now happen that
$\phi_{i_1}=\phi_{i_2}$ for some $i_1\neq i_2$. We therefore adjust the definition
of $p$ and $q$ to $p=\lcm(x-\phi_1,\dots,x-\phi_n)$, $q=\lcm(x-\psi_1,\dots,x-\psi_m)$.
Then $p$ and $q$ are squarefree and for the set of roots of $p\otimes q$
we obtain
\[
\{\,\phi_i\psi_j:i=1,\dots,n;j=1,\dots,m\,\}=\{\rho_1,\dots,\rho_\ell\},
\]
as desired.

\begin{example}
  \begin{enumerate}
  \item To find the factorization $(x-\phi_1^2)(x-\phi_1\phi_2)(x-\phi_2^2)=
    (x-\phi_1)(x-\phi_2)\otimes(x-\phi_1)(x-\phi_2)$,
    set $\rho_1=\phi_1^2$, $\rho_2=\phi_1\phi_2$, $\rho_3=\phi_2^2$. Then
    a suitable choice for $\pi\colon\{1,2\}\times\{1,2\}\to\{1,2,3\}$ is
    given by
    \begin{center}
      \begin{tabular}{c|cc}
        $\pi$ & $1$ & $2$ \\\hline
        $1$ & $1$ & $2$ \\
        $2$ & $2$ & $3$ 
      \end{tabular}
    \end{center}
    because
    \[
       \frac{\rho_{\pi(1,1)}}{\rho_{\pi(1,2)}}
       = \frac{\rho_1}{\rho_2} = \frac{\phi_1}{\phi_2}
       = \frac{\rho_2}{\rho_3}
       = \frac{\rho_{\pi(2,1)}}{\rho_{\pi(2,2)}}
    \]
    and
    \[
    \frac{\rho_{\pi(1,1)}}{\rho_{\pi(2,1)}}
    = \frac{\rho_1}{\rho_2} = \frac{\phi_1}{\phi_2}
    = \frac{\rho_2}{\rho_3}
    = \frac{\rho_{\pi(1,2)}}{\rho_{\pi(2,2)}}.
    \]
  \item
    Consider $r=(x-\tfrac12)(x-\tfrac14)(x-1)(x-2)$, i.e.,
    $\rho_1=\frac12$, $\rho_2=\frac14$, $\rho_3=1$, $\rho_4=2$. A possible
    choice for $\pi\colon\{1,2\}\times\{1,2,3\}\to\{1,2,3,4\}$ is
    \begin{center}
      \begin{tabular}{c|ccc}
        $\pi$ & $1$ & $2$ & $3$ \\\hline
          $1$ & $1$ & $3$ & $4$ \\
          $2$ & $2$ & $1$ & $3$
      \end{tabular}
    \end{center}
    because
    \begin{alignat*}1
      &\Bigl\{
      \frac{\rho_{\pi(1,1)}}{\rho_{\pi(1,2)}},
      \frac{\rho_{\pi(2,1)}}{\rho_{\pi(2,2)}}
      \Bigr\}=\Bigl\{
      \frac{\rho_1}{\rho_3},
      \frac{\rho_2}{\rho_1}
      \Bigr\}=\bigl\{\frac12\bigr\}\\
      &\Bigl\{
      \frac{\rho_{\pi(1,1)}}{\rho_{\pi(1,3)}},
      \frac{\rho_{\pi(2,1)}}{\rho_{\pi(2,3)}}
      \Bigr\}=\Bigl\{
      \frac{\rho_1}{\rho_4},
      \frac{\rho_2}{\rho_3}
      \Bigr\}=\bigl\{\frac14\bigr\}\\
      &\Bigl\{
      \frac{\rho_{\pi(1,2)}}{\rho_{\pi(1,3)}},
      \frac{\rho_{\pi(2,2)}}{\rho_{\pi(2,3)}}
      \Bigr\}=\Bigl\{
      \frac{\rho_3}{\rho_4},
      \frac{\rho_1}{\rho_3}
      \Bigr\}=\bigl\{\frac12\bigr\}
    \end{alignat*}
    and
    \[
    \Bigl\{
        \frac{\rho_{\pi(1,1)}}{\rho_{\pi(2,1)}},
        \frac{\rho_{\pi(1,2)}}{\rho_{\pi(2,2)}},
        \frac{\rho_{\pi(1,3)}}{\rho_{\pi(2,3)}}
        \Bigr\}=\Bigl\{
        \frac{\rho_1}{\rho_2},
        \frac{\rho_3}{\rho_1},
        \frac{\rho_4}{\rho_3}
        \Bigr\}=\bigl\{2\bigr\}
    \]
  \end{enumerate}
\end{example}

\section{Searching for Assignments}\label{sec:search}

We now turn to the question how for a given $r=(x-\rho_1)\cdots(x-\rho_\ell)\in
k[x]$ we can find a map $\pi$ as required. Of course, since $\ell$ is finite,
there are only finitely many possible choices for $n$ and $m$ such that
$n+m\leq\ell\leq nm$, and for each choice $n,m$ there are only finitely many
functions $\pi\colon\{1,\dots,n\}\times\{1,\dots,m\}\to\{1,\dots,\ell\}$. We can
simply try them all. But going through all these $(nm)^\ell$ many functions one
by one would take very long.

In order to improve the efficiency of the search, we can exploit the fact that
for most partial functions $\pi$ it is easy to see that they cannot be extended
to a total function with the required properties. We can further reduce the
search space by taking into account that the order of the roots of the factors
is irrelevant, i.e., we can restrict the search to functions $\pi$ with
$\pi(1,1)\leq\pi(2,1)\leq\cdots\leq\pi(n,1)$ and
$\pi(1,1)\leq\pi(1,2)\leq\cdots\leq\pi(1,m)$. Furthermore, because of
surjectivity, the root $\rho_1$ must be reached, and we can choose to set
$\pi(1,1)=1$ without loss of generality. Next, discard all functions with
$\pi(i,j_1)=\pi(i,j_2)$ for some $i,j_1,j_2$ with $j_1\neq j_2$ or with
$\pi(i_1,j)=\pi(i_2,j)$ for some $i_1,i_2,j$ with $i_1\neq i_2$, because these
just signal some roots of a factor of $r$ several times without providing any
additional information. So we can in fact enforce
$1=\pi(1,1)<\pi(2,1)<\cdots<\pi(n,1)$ and $\pi(1,1)<\pi(1,2)<\cdots<\pi(1,m)$.
Next, $\pi$ is a solution iff
$\pi^\top\colon\{1,\dots,m\}\times\{1,\dots,n\}\to\{1,\dots,\ell\}$ with
$\pi^\top(i,j)=\pi(j,i)$ is a solution. We can therefore restrict the search to
functions where $n\leq m$. 

The following algorithm takes these observations into account. It maintains an
assignment table $M$ which encodes a function~$\pi\colon
\{1,\dots,n\}\times\{1,\dots,m\}\to\{1,\dots,\ell\}$ with
\[
\frac{\rho_{\pi(1,j_1)}}{\rho_{\pi(1,j_2)}}=\frac{\rho_{\pi(2,j_1)}}{\rho_{\pi(2,j_2)}}=\cdots=\frac{\rho_{\pi(n,j_1)}}{\rho_{\pi(n,j_2)}}
\]
for all $i,j_1,j_2$ and
\[
\frac{\rho_{\pi(i_1,1)}}{\rho_{\pi(i_2,1)}}=\frac{\rho_{\pi(i_1,2)}}{\rho_{\pi(i_2,2)}}=\cdots=\frac{\rho_{\pi(i_1,m)}}{\rho_{\pi(i_2,m)}}.
\]
for all $i_1,i_2,j$. At every recursion level, the candidate under
consideration is extended to a function $\pi$ with $\pi(n+1,1)=p$ for
some~$p$. As soon as $p$ is chosen, there is for each $j=2,\dots,m$ at most one
choice $q\in\{1,\dots,\ell\}$ for the value of $\pi(n+1,j)$. The matrix $M$
stores these values $q$ and marks the indices $j$ for which no $q$ exists with
$q=0$. The result is a function $\{1,\dots,n+1\}\times\{1,\dots,\tilde
m\}\to\{1,\dots,\ell\}$ for some $\tilde m\leq m$.  If this function is
surjective, we have found a solution.  Otherwise, we proceed recursively unless
we already have $n+1=\tilde m$, because in this case any further extension could
only produce transposes of solutions that will be found at some other stage of
the search.

\medskip\noindent
INPUT: The roots $\rho_1,\dots,\rho_\ell$ of some square-free polynomial $r\in k[x]$.\\
OUTPUT: A list of functions $\pi$ as required for solving the factorization problem.

\smallskip
\step 10 let $M=((M[i,j]))_{i,j=1}^\ell$ be a matrix with $M[1,j]=j$ for $j=1,\dots,\ell$.
\step 20 call the procedure $\mathrm{addRow}(M, 2)$ as defined below.
\step 30 stop.

\smallskip
\step 40 procedure $\mathrm{addRow}(M, n)$ 
\step 51 for $p=M[n-1,1]+1,\dots,\ell$ do:
\step 62 set the $n$th row of $M$ to $(p,0,\dots,0)$ and let $J$ be the empty list
\step 72 for $j=2,\dots,\ell$ do:
\step 83 if $M[n-1,j]\neq0$ and there exists $q\in\{1,\dots,\ell\}$ such that $\rho_1/\rho_p=\rho_j/\rho_q$ and $\rho_1/\rho_j=\rho_p/\rho_q$
\step 94 set $M[n,j] = q$ and append $j$ to $J$
\step {10}2 if $\{M[i,j]:i=1,\dots,n; j\in J\}=\{1,\dots,\ell\}$ then:
\step {11}3 report the solution $\pi\colon\{1,\dots,n\}\times\{1,\dots,|J|\}\to\{1,\dots,\ell\}$ with $\pi(i,j)=M[i,J[j]]$ for all $i,j$.
\step {12}2 else if $|\{j:M[n,j]\neq0\}|<n$ then
\step {13}3 recursively call the procedure $\mathrm{addRow}(M, n+1)$

\medskip
In the interest of readability, we have refrained from some obvious optimizations.
For example, an actual implementation might perform some precomputation in order to
improve the search for $q$ in Step~8.

It is not hard to implement the algorithm. A Mathematica implementation by the
authors is available on the website of this paper,
\url{http://www.math.rutgers.edu/~zeilberg/mamarim/mamarimhtml/Cfac.html}.  The
relevant function is \texttt{CFiniteFactor}.

\begin{example}
  Let $r=(x-\rho_1)\cdots(x-\rho_6)$ where
  $\rho_1=-8$, $\rho_2=-6$, $\rho_3=-4$, $\rho_4=-3$,
  $\rho_5=-2$, $\rho_6=-1$.

  After initialisation, at the first level of the recursion, there are five choices
  for the first entry in the second row of~$M$. Each of them uniquely determines the
  rest of the row, as follows (writing $\cdot$ for~$0$):

  \begin{alignat*}1
    &\begin{pmatrix}
    1&2&3&4&5&6\\
    2&\cdot&4&\cdot&\cdot&\cdot
    \end{pmatrix},\\
    &\begin{pmatrix}
    1&2&3&4&5&6\\
    3&4&5&\cdot&6&\cdot
    \end{pmatrix},\\
    &\begin{pmatrix}
    1&2&3&4&5&6\\
    4&\cdot&\cdot&\cdot&\cdot&\cdot
    \end{pmatrix},\\
    &\begin{pmatrix}
    1&2&3&4&5&6\\
    5&\cdot&6&\cdot&\cdot&\cdot
    \end{pmatrix},\\
    &\begin{pmatrix}
    1&2&3&4&5&6\\
    6&\cdot&\cdot&\cdot&\cdot&\cdot
    \end{pmatrix}.
  \end{alignat*}
  The second of these matrices corresponds to a solution
  \[
    \pi\colon\{1,2\}\times\{1,2,3,4\}\to\{1,2,3,4,5,6\},
  \]
  which gives rise to the factorization
  \[
  r=(x-1)(x-\tfrac12)\otimes(x+8)(x+6)(x+4)(x+2),
  \]
  while the other partial solutions cannot be continued to further solutions.   
\end{example}

\section{Multiple Roots}

Let us now drop the condition that $r\in k[x]$ is square free. Write $r^\ast$
for the square free part of~$r$. It is clear from equation~\eqref{eq:1} that
when $p,q\in k[x]$ are such that $r=p\otimes q$, then $r^\ast=p^\ast\otimes
q^\ast$, where $p^\ast,q^\ast$ denote the square free parts of $p$ and~$q$,
respectively. It is therefore natural to first determine factorizations of the
square free part $r^\ast$ of~$r$ and in a second step obtain $p$ and $q$ from
$p^\ast$ and $q^\ast$ (if possible) by assigning appropriate multiplicities to
their roots. As the multiplicities in $p$ or $q$ cannot exceed those in~$r$,
there are again just finitely many candidates and we could simply try them all.
And again, the search can be improved because many possibilities can be ruled
out easily. In fact, the freedom for the multiplicities is so limited that we
can compute them rather than search for them.

First consider the case when $p^\ast$ and $q^\ast$ were obtained from an injective
map $\pi$, i.e., the case when there are no product clashes. In this case, each
root $\rho_\ell$ of $r^\ast$ corresponds to exactly one product $\phi_i\psi_j$
of a root $\phi_i$ of $p^\ast$ and a root $\psi_j$ of~$q^\ast$. The
multiplicities $e_i$ of $\phi_i$ in $p$ and $\epsilon_j$ of $\psi_j$ in~$q$,
respectively, must be such that $e_i+\epsilon_j-1$ equals the
multiplicity of $\rho_\ell$ in~$r$. This gives a linear system of equations.
Every solution of this system in the positive integers gives rise to a
factorization for~$r$, and if there is no solution for the linear system of
any of the factorizations of the square-free part~$r^\ast$, then $r$ admits no factorization.

When there are product clashes, there are roots $\rho$ of $r$ which are obtained
in several distinct ways as products of roots of $p$ and~$q$, for instance
$\rho=\phi_{i_1}\psi_{j_1}=\phi_{i_2}\psi_{j_2}$ for some
$(i_1,j_1)\neq(i_2,j_2)$.  If $m$ is the multiplicity of $\rho$ in~$r$, then the
requirement for the multiplicities
$e_{i_1},e_{i_2},\epsilon_{j_1},\epsilon_{j_2}$ of
$\phi_{i_1},\phi_{i_2},\psi_{j_1},\psi_{j_2}$ in $p$ and~$q$, respectively, is
that
\[
\max(e_{i_1}+\epsilon_{j_1}-1,e_{i_2}+\epsilon_{j_2}-1)=m.
\]
We obtain a
system of such equations, one equation for reach root of~$r$.  Such systems are
known as tropical linear systems, and algorithms are known for finding their
solutions in polynomial time~\cite{grigoriev13}.

\begin{example}
  \begin{enumerate}
  \item 
  Let $r=(x-2)(x+2)^2(x-3)^2(x+3)^3$.
  We have seen earlier that the square free part $r^\ast$ of $r$ admits
  two distinct factorizations
  \begin{alignat*}1
    r^\ast&=(x-1)(x+1)\otimes(x-2)(x+3)\\
         &=(x-1)(x+1)\otimes(x-2)(x-3).
  \end{alignat*}
  Assigning multiplicities to the first, we get
  \begin{alignat*}1
    &(x-1)^{e_1}(x+1)^{e_2}\otimes(x-2)^{\epsilon_1}(x+3)^{\epsilon_2}\\
    &=(x{+}2)^{e_1+\epsilon_1-1}(x{-}3)^{e_1+\epsilon_2-1}(x{-}2)^{e_2+\epsilon_1-1}(x{+}3)^{e_2+\epsilon_2-1}.
  \end{alignat*}
  Comparing the exponents to those of~$r$ gives the linear system
  \begin{alignat*}3
     e_1+\epsilon_1-1&=2,\qquad
    &e_1+\epsilon_2-1&=2,\\
     e_2+\epsilon_1-1&=1,\qquad
    &e_2+\epsilon_2-1&=3,
  \end{alignat*}
  which has no solution. For the second factorization, we get
  \begin{alignat*}1
    &(x-1)^{e_1}(x+1)^{e_2}\otimes(x-2)^{\epsilon_1}(x-3)^{\epsilon_2}\\
    &=(x{+}2)^{e_1+\epsilon_1-1}(x{+}3)^{e_1+\epsilon_2-1}(x{-}2)^{e_2+\epsilon_1-1}(x{-}3)^{e_2+\epsilon_2-1}.
  \end{alignat*}
  Comparing the exponents to those of~$r$ gives the linear system
  \begin{alignat*}3
    e_1+\epsilon_1-1&=2,\qquad
    &e_1+\epsilon_2-1&=3,\\
    e_2+\epsilon_1-1&=1,\qquad
    &e_2+\epsilon_2-1&=2,
  \end{alignat*}
  whose unique solution in the positive integers is $e_1=2$, $e_2=1$, $\epsilon_1=1$, $\epsilon_2=2$,
  thus
  \[
    r = (x-1)^2(x+1)\otimes(x-2)(x-3)^2.
  \]
  \item
    Let $r=(x-\tfrac12)^2(x-\tfrac14)(x-1)^2(x-2)^3$.
    We have seen earlier that the square free part $r^\ast$ of $r$ admits the factorization
    \[
      r^\ast= (x-\tfrac12)(x-\tfrac14)\otimes(x-1)(x-2)(x-4).
    \]
    Assigning multiplicities to the factors, we get
    \begin{alignat*}1
      &(x-\tfrac12)^{e_1}(x-\tfrac14)^{e_2}\otimes(x-1)^{\epsilon_1}(x-2)^{\epsilon_2}(x-4)^{\epsilon_3}\\
      &=(x-\tfrac12)^{\max(e_1+\epsilon_1-1,e_2+\epsilon_2-1)}\\
      &\hphantom{{}={}}(x-1)^{\max(e_1+\epsilon_2-1,e_2+\epsilon_3-1)}\\
      &\hphantom{{}={}}(x-2)^{e_1+\epsilon_3-1}(x-\tfrac14)^{e_2+\epsilon_1-1}.
    \end{alignat*}
    Comparing the exponents to the exponents of the factors of~$r$ gives a tropical linear system
    in the unknowns $e_1,e_2,\epsilon_1,\epsilon_2,\epsilon_3$, which turns out to have two solutions. They
    correspond to the two factorizations
    \begin{alignat*}1
      r&=(x-\tfrac12)^2(x-\tfrac14)\otimes(x-1)(x-2)(x-4)^2\\
       &=(x-\tfrac12)^2(x-\tfrac14)\otimes(x-1)(x-2)^2(x-4)^2
    \end{alignat*}    
  \end{enumerate}  
\end{example}

\section{When we don't want to find the roots}

Sometimes our polynomials are with integer coefficients, and we prefer not to factorize them over the complex numbers. Of course, all
the roots are algebraic numbers, by definition, and computer-algebra systems know how to compute with them (without ``cheating'' and using
floating-point approximations), but it may be more convenient to find the tensor product (in the generic case: no product
clashes and no repeated roots) of $p=p_0+\cdots+p_mx^m$ and $q=q_0+\cdots+q_nx^n$,
a certain polynomial $r$ of degree~$mn$, as follows.
If the roots of $p$ are $\phi_1,\dots,\phi_n$ and  the roots of $q$ are $\psi_1,\dots,\psi_m$, then
the roots of $p\otimes q$ are, of course
$$
\{ \, \phi_i \psi_j \mid 1 \leq i \leq m, \, 1 \leq j \leq n \, \}.
$$
Let $P_k(p):=\sum_{i=1}^m \phi_i^k$ be the {\it power-sum symmetric functions}~\cite{macdonald95}, then of course
$$
 P_k(p\otimes q)= P_k(p) P_k(q), \quad 1 \leq k \leq nm.
$$
Now using {\it Newton's relations} (e.g. \cite{macdonald95}, Eq. I.(2.11') p.~23), one can go back and forth from
the elementary symmetric functions (essentially the coefficients of the polynomial up to sign) to the power-functions,
and {\it back,} enabling us easily to compute the tensor product without factorizing.

If you define the reverse of a polynomial~$p$, to be $p^{*}(x):=x^d p(1/x)$, where $d$ is the degree of~$p$,
then $p\otimes p^{*}$ has, of course, the factor $(x-1)^d$ but otherwise (generically) all distinct roots,
unless it has good reasons not to.
On the other hand, if $r=p \otimes q$ for some non-trivial polynomials $p$ and $q$ then  $r \otimes r^{*}$ has repeated roots,
and the {\it repetition profile} can be easily predicted as above, or ``experimentally''. So using
this approach it is easy to {\it test} quickly whether $r$ ``factorizes'', in the tensor-product sense.
However, to actually find the factors would take more effort.

This is implemented in the Maple package accompanying this article, linked to from
\url{http://www.math.rutgers.edu/~zeilberg/mamarim/mamarimhtml/Cfac.html}.
The tensor product operation is procedure {\tt Mul} and the testing procedure is {\tt TestFact}.

\section{Linear Combinations of Factorizations}

For almost all polynomials $r\in k[x]$ there does not exist a factorization.
When no factorization exists, we may wonder whether $r$ admits a decomposition of
a more general type. For example, we can ask whether there exist polynomials
$p_1,p_2,q_1,q_2$ of degree at least two such that
\[
  r = \lcm(p_1\otimes q_1, \ p_2\otimes q_2).
\]
Translated to the language of C-finite sequences, this means that we seek to write
a given C-finite sequence $(a_n)_{n=0}^\infty$ as
\[
  a_n = b_nc_n + u_nv_n
\]
for C-finite sequences $(b_n)_{n=0}^\infty$, $(c_n)_{n=0}^\infty$,
$(u_n)_{n=0}^\infty$, $(v_n)_{n=0}^\infty$, none of which should satisfy a first-order
recurrence in order to make the problem nontrivial.

\def\im{\operatorname{im}}
It is not difficult to adapt the algorithm in Section~\ref{sec:search}
so that it can also discover such factorizations.
Suppose that $r$ is squarefree. Then, instead of searching for 
a single surjective map
\[\pi\colon\{1,\dots,n\}\times\{1,\dots,m\}\to\{1,\dots,\ell\},\]
it suffices to find two functions
\begin{alignat*}1
  &\pi_1\colon\{1,\dots,n_1\}\times\{1,\dots,m_1\}\to\{1,\dots,\ell\}\\
  &\pi_2\colon\{1,\dots,n_2\}\times\{1,\dots,m_2\}\to\{1,\dots,\ell\}
\end{alignat*}
satisfying the same conditions previously requested for $\pi$ but with surjectivity replaced by
$\im\pi_1\cup\im\pi_2=\{1,\dots,\ell\}$. Once two such maps $\pi_1,\pi_2$ have been found,
we can construct $p_1,p_2,q_1,q_2$ by choosing $\phi_1^1$ and $\phi_1^2$ arbitrarily,
setting $\psi_1^1=\rho_{\pi_1(1,1)}/\phi_1^1$, $\psi_1^2=\rho_{\pi_2(1,1)}/\phi_1^2$ and
\begin{alignat*}3
  \phi_i^1 &= \phi_1^1\frac{\rho_{\pi_1(i,1)}}{\rho_{\pi_1(1,1)}}, &\qquad
  \psi_j^1 &= \psi_1^1\frac{\rho_{\pi_1(1,j)}}{\rho_{\pi_1(1,1)}}, \\
  \phi_i^2 &= \phi_1^2\frac{\rho_{\pi_2(i,1)}}{\rho_{\pi_2(1,1)}}, &\qquad
  \psi_j^2 &= \psi_1^2\frac{\rho_{\pi_2(1,j)}}{\rho_{\pi_2(1,1)}}
\end{alignat*}
for all $i,j$ in question. Then $p_1:=\prod_{i=1}^{n_1}(x-\phi_i^1)$, $q_1:=\prod_{i=1}^{m_1}(x-\psi_j^1)$,
$p_2:=\prod_{i=1}^{n_2}(x-\phi_i^2)$, $q_2:=\prod_{i=1}^{m_2}(x-\psi_j^2)$,
are such that $r=\lcm(p_1\otimes q_1,p_2\otimes q_2)$. 

In order to search for a pair $\pi_1,\pi_2$, we can search for $\pi_1$ very much like we searched
for $\pi$ before, and for each partial solution encountered during the recursion, initiate a search
for another function $\pi_2$ which is required to hit all the indices $1,\dots,\ell$ not hit by
the partial solution~$\pi_1$. Note that it is fine if some indices are hit by both $\pi_1$ and~$\pi_2$.
The suggested modification amounts to replacing lines 12 and~13 of the algorithm from Section~\ref{sec:search}
by the following:

\medskip
\step {12}2 else
\step {13}3 let $Q=\{M[i,j]:i=1,\dots,n;j\in J\}$.
\step {14}3 let $M_2$ be an $\ell\times\ell$-matrix with $(1,\dots,\ell)$ as first row. 
\step {15}3 call the procedure $\mathrm{addRow}_2(M_2,2,Q)$ defined below.
\step {16}3 for each function $\pi_2$ it reports, report $(\pi,\pi_2)$.
\step {17}3 if no $\pi_2$ is found and $|\{j:M[n,j]\neq0\}|<n$ then
\step {18}4 recursively call $\mathrm{addRow}(M,n+1)$

\medskip
\step {19}0 procedure $\mathrm{addRow}_2(M,n,Q)$
\step {20}1 [lines 5--9 literally as in the definition of $\mathrm{addRow}$]
\step {21}2 if $\{1,\dots,\ell\}\setminus Q\subseteq\{M[i,j]:i=1,\dots,n;j\in J\}$ then:
\step {22}3 [line 11 literally as in the definition of $\mathrm{addRow}$]
\step {23}2 else if $|\{j:M[n,j]\neq0\}|<n$ then
\step {24}3 recursively call $\mathrm{addRow}_2(M,n+1,Q)$. 

\medskip

This settles the case of square free input. The extension to arbitrary
polynomials is like in the previous section. For every factorization of the
square free part we can assign variables for the multiplicities of all the roots
and compare the resulting multiplicities for $\lcm(p_1\otimes q_1,p_2\otimes
q_2)$ to those of~$r$. This gives again a tropical linear system of equations
which can be solved with Grigoriev's algorithm~\cite{grigoriev13}.

\begin{example}
  The polynomial $r=(x-1)(x-2)(x-3)(x-4)(x-6)(x-12)$ cannot be written as
  $r=p\otimes q$ for some $p,q\in k[x]$. However, we have the representation
  \[
    r=\lcm(p_1\otimes q_1,p_2\otimes q_2)
  \]
  for
  \begin{alignat*}3
    p_1 &= (x-1)(x-2),&\qquad
    p_2 &= (x-1)(x-3),\\
    q_1 &= (x-2)(x-3),&\qquad
    q_2 &= (x-1)(x-4).
  \end{alignat*}
  Note that the roots $3$ and $4$ of $r$ are produced by both $p_1\otimes q_1$
  and $p_2\otimes q_2$.  
\end{example}





\section{Examples}

Our main motivation for studying the factorization problem for C-finite sequences
are two interesting identities that can be interpreted as such factorizations. They
both originate from the transfer matrix method.

\def\-{{\tikz[x=1.5pt,y=1.5pt]\draw[very thin](0,0) rectangle (1,1) (1,0) rectangle (2,1);}}
\def\|{{\tikz[x=1.5pt,y=1.5pt]\draw[very thin](0,0) rectangle (1,1) (0,1) rectangle (1,2);}}

The first is a tiling problem studied in~\cite{kasteleyn61,fisher61}, and more recently in~\cite{zeilberger06a}.
Given a rectangle of size $m\times n$, the question is
in how many different ways we can fill it using tiles of size $2\times 1$ or $1\times 2$.
If $n$ and $m$ are even, it turns out that
\[
 T_{n,m} = 2^{nm/2}\prod_{i=1}^{m/2}\prod_{j=1}^{n/2}\biggl(z_\-^2\cos^2\Bigl(\frac{i\pi}{m+1}\Bigr)+z_\|^2\cos^2\Bigl(\frac{j\pi}{n+1}\Bigr)\biggr)
\]
is a bivariate polynomial in the variables $z_\-,z_\|$ where the coefficient of a monomial $z_\-^uz_\|^v$ is exactly
the number of tilings of the $m\times n$ rectangle that uses exactly $u$~tiles of size $2\times 1$
and $v$~tiles of size $1\times 2$. 
The transfer matrix method can be used to prove this result automatically for every fixed
$m$ and arbitrary $n$ (or vice versa). For every fixed choice of $m$ (say), it delivers
a polynomial~$r$ which encodes a recurrence for $(T_{n,m})_{n=0}^\infty$. For every fixed $i\in\{1,\dots,m\}$,
the sequence
\begin{alignat*}1
  &2^{n/2}\prod_{j=1}^n\biggl(z_\-^2\cos^2\Bigl(\frac{i\pi}{m+1}\Bigr)+z_\|^2\cos^2\Bigl(\frac{j\pi}{n+1}\Bigr)\biggr)\\
  &{}= \frac1wz_\|^n T_n(\sqrt w) + \Bigl(1-\frac1w\Bigr)z_\|^n U_n(\sqrt w)
\end{alignat*}
with $w=1+\bigl(\frac{z_\-}{z_\|}\cos(\frac{i\pi}{m+1})\bigr)^2$ and $T_n$ and $U_n$ the Chebyshev polynomials
of the first and second kind, 
is C-finite with respect to~$n$. An annihilating polynomial is
\[
 p_i = x^2 - 2 \Bigl(z_\|^2 + 2 z_\-^2\cos^2\bigl(\tfrac{i\pi}{2m+1}\bigr)\Bigr)x + z_\|^4. 
\]
The formula for $T_{n,m}$ can be proven for each particular choice of $m$ and arbitrary $n$ by checking
$r=p_1\otimes\cdots\otimes p_m$ and comparing the first $2^m$ initial terms.
While the standard algorithms can confirm the correctness of some conjectured
factorization, the algorithm described in the present paper can
help discover the factorization in the first place, taking only $r$ as input.
Fisher, Temperly~\cite{fisher61} or Kasteleyn~\cite{kasteleyn61} would probably
have found it useful back in the 1960s to apply the algorithm to $m=2,4,6,8,10$
and to detect the general pattern from the outputs.

The second identity has a similar nature. It describes the Ising model on an
$n\times m$ grid wrapped around a torus~\cite{onsager44,thompson72}. Starting
from a certain model in statistical physics that we do not want to explain here,
the transfer matrix method produces for every fixed $m\in\set N$ an annihilating
polynomial~$r$ of degree~$2^m$ for a certain C-finite sequence in~$n$. The
asymptotic behaviour of this sequence for $n\to\infty$ is of interest. In view
of Theorem~\ref{thm:main}, it is goverend by the root of~$r$ with the largest
absolute value. Onsager discovered that this largest root of~$r$ is equal to
\[
 (2\sinh(2\nu))^{m/2}\exp\bigl(\tfrac12(\gamma_1+\gamma_3+\cdots+\gamma_{2m-1})\bigr)
\]
where $\nu$ is some physical constant and $\gamma_k$ is defined as
\def\arccosh{\operatorname{arccosh}}
\[
  \gamma_k = \arccosh\bigl(\cosh(2\nu)\coth(2\nu) - \cos(\tfrac{\pi k}m)\bigr)
\]
for $k=1,3,\dots,2m-1$ (compare eq.~(V.5.1) (p. 131) in~\cite{thompson72}). 

Let us translate these formulas to a more familiar form.
First note that because of periodicity and symmetry of the cosine,
we have $\gamma_k=\gamma_{2m-k}$ for $k=1,3,\dots$. Hence each of the
$\gamma_k$ in the argument of $\exp$ appears twice, except the
middle term $\gamma_m$, which only appears for odd~$m$.
Set $z=\exp(\nu)$ and $x_k=\exp(\gamma_k)$ for $k=1,3,\dots,2m-1$.
Then $2\sinh(2\nu)=z^2-z^{-2}$, and Onsager's expression for the
largest root of $r$ simplifies to
\[
  \left\{\begin{array}{ll}
  (z^2+z^{-2})^{m/2}x_1x_3\cdots x_{m-1} &\text{if $m$ is even}\\
  (z^2+z^{-2})^{(m-1)/2}(1+z^2)x_1x_3\cdots x_{m-1}&\text{if $m$ is odd}.
  \end{array}\right.
\]
For the second case we have used $\sqrt{(z^2+z^{-2})x_m}=1+z^2$.
The equation for $\gamma_k$ says that $x_k$ is a root of 
\[
 p_k:=x^2 + \Bigl(2\cos(\tfrac{\pi k}m) -\frac{(z^4+1)^2}{(z^4-1)z^2}\Bigr)x + 1.
\]
Set $q=x-(z^2-z^{-2})^{m/2}$ when $m$ is even and set $q=x-(z^2-z^{-2})^{(m-1)/2}(1+z^2)$
when $m$ is odd. Then Onsager's formula says that the largest root of $r$ is equal
to the largest root of $q\otimes p_1\otimes p_3\otimes\cdots\otimes p_{m-1}$.

In fact, the polynomial $q\otimes p_1\otimes p_3\otimes\cdots\otimes p_{m-1}\in\set Q(z)[x]$ happens to be
exactly the irreducible factor of $r\in\set Q(z)[x]$ corresponding to the largest root of~$r$. Therefore,
our algorithm applied to this irreducible factor of $r$ could have helped Onsager discover his formula.

\bibliographystyle{plain}
\bibliography{bib}

\begin{thebibliography}{10}

\bibitem{cha14}
Yongjae Cha.
\newblock Closed form solutions of linear difference equations in terms of
  symmetric products.
\newblock {\em Journal of Symbolic Computation}, 60:62--77, 2014.

\bibitem{fisher61}
M.~Fisher and H.~Temperley.
\newblock Dimer problems in statistical mechanics--an exact result.
\newblock {\em Philos. Mag.}, 6:1061--1063, 1961.

\bibitem{grigoriev13}
Dima Grigoriev.
\newblock Complexity of solving tropical linear systems.
\newblock {\em Computational Complexity}, 22:71--88, 2013.

\bibitem{hessinger97}
Sabrina Hessinger.
\newblock {\em Computing Galois Groups of Linear Differential Equations of
  Order Four}.
\newblock PhD thesis, North Carolina State University, 1997.

\bibitem{kasteleyn61}
P.~W. Kasteleyn.
\newblock The statistics of dimers on a lattice: {I.} the number of dimer
  arrangements in a quadratic lattice.
\newblock {\em Physica}, 27:1209--1225, 1961.

\bibitem{kauers10j}
Manuel Kauers and Peter Paule.
\newblock {\em The Concrete Tetrahedron}.
\newblock Springer, 2011.

\bibitem{macdonald95}
Ian Macdonald.
\newblock {\em Symmetric Functions and Hall Polynomials}.
\newblock Clarendon Press, Oxford, 2nd edition, 1995.

\bibitem{onsager44}
Lars Onsager.
\newblock Crystal statistics, {I.} a two-dimensional model with an
  order-disorder transition.
\newblock {\em Physical Review}, 65:117--149, 1944.

\bibitem{singer85a}
Michael~F. Singer.
\newblock Solving homogeneous linear differential equations in terms of second
  order linear differential equations.
\newblock {\em American Journal of Mathematics}, 107(3):663--696, 1985.

\bibitem{stanley99}
Richard~P. Stanley.
\newblock {\em Enumerative Combinatorics, Volume 2}.
\newblock Cambridge Studies in Advanced Mathematics 62. Cambridge University
  Press, 1999.

\bibitem{thompson72}
Colin~J. Thompson.
\newblock {\em Mathematical Statistical Mechanics}.
\newblock Princeton University Press, 1972.

\bibitem{hoeij02a}
Mark van Hoeij.
\newblock Decomposing a 4th order linear differential equation as a symmetric
  product.
\newblock {\em Banach Center Publications}, 58:89--96, 2002.

\bibitem{hoeij07}
Mark van Hoeij.
\newblock Solving third order lienar differential equations in terms of second
  order equations.
\newblock In {\em Proceedings of ISSAC'07}, pages 355--360, 2007.

\bibitem{zeilberger90}
Doron Zeilberger.
\newblock A holonomic systems approach to special function identities.
\newblock {\em Journal of Computational and Applied Mathematics}, 32:321--368,
  1990.

\bibitem{zeilberger06a}
Doron Zeilberger.
\newblock {CounTilings}.
\newblock The Personal Journal of Shalosh B. Ekhad and Doron Zeilberger, 2006.

\bibitem{zeilberger13}
Doron Zeilberger.
\newblock The {C}-finite ansatz.
\newblock {\em The Ramanujan Journal}, 31(1):23--32, 2013.

\end{thebibliography}

\end{document}